\documentclass[12pt,preprint]{aastex}

\usepackage{graphicx, color}
\usepackage{dcolumn}
\usepackage{bm}
\usepackage{booktabs}
\usepackage{natbib}
\usepackage{lscape}

\newcommand {\dif}[3][]{\frac{d^{#1}#2}{d#3^{#1}}}

\newcommand {\gsim}{\hspace{0.3em}\raisebox{0.4ex}{$>$}\hspace{-0.75em}\raisebox{-.7ex}{$\sim$}\hspace{0.3em}}

\makeatletter
\def\mart{\@ifnextchar[{\mart@@}{\mart@}}
\def\mart@@[#1]#2{\sqrt[#1]{\mathstrut{#2}}}
\def\mart@#1{\sqrt{\mathstrut{#1}}}
\makeatother
\newcommand {\Alfven}{Alfv\'{e}n}
\newcommand{\myemail}{minoshim@stelab.nagoya-u.ac.jp}

\newcommand{\SOHO}{\it SOHO}
\newcommand{\TRACE}{\it TRACE}
\newcommand{\RHESSI}{\it RHESSI}
\newcommand{\Hinode}{\it Hinode}

\newcommand{\gyros}{gyrosynchrotron}

\begin{document}

\shorttitle{Electron Acceleration in the 2006 December 13 Flare}
\shortauthors{Minoshima et al.}

\title{Multi-Wavelength Observation of Electron Acceleration in the 2006 December 13 Flare}

\author{T. Minoshima\altaffilmark{1,2}, S. Imada\altaffilmark{3}, T. Morimoto\altaffilmark{4}, T. Kawate\altaffilmark{5}, H. Koshiishi\altaffilmark{6}, M. Kubo\altaffilmark{7}, S. Inoue\altaffilmark{8}, H. Isobe\altaffilmark{5}, S. Masuda\altaffilmark{2}, S. Krucker\altaffilmark{9}, and T. Yokoyama\altaffilmark{1}}
\altaffiltext{1}{
Department of Earth and Planetary Science, Graduate School of Science, University of Tokyo,
7-3-1 Hongo, Bunkyo-ku, Tokyo 113-0033, Japan
}
\altaffiltext{2}{
Solar-Terrestrial Environment Laboratory, Nagoya University,
Furo-cho, Chikusa-ku, Nagoya 464-8601, Japan
}
\altaffiltext{3}{
National Astronomical Observatory of Japan, 2-21-1 Osawa, Mitaka, Tokyo 181-8588, Japan
}
\altaffiltext{4}{
Ibaraki University, 2-1-1 Bunkyo, Mito, Ibaraki 310-8512, Japan
}
\altaffiltext{5}{
Kwasan and Hida Observatories, Kyoto University, Yamashina-ku, Kyoto 607-8471, Japan 
}
\altaffiltext{6}{
Tsukuba Space Center, Japan Aerospace Exploration Agency, 2-1-1 Sengen, Tsukuba, Ibaraki 305-8505, Japan 
}
\altaffiltext{7}{
High Altitude Observatory, National Center for Atmospheric Research, P O Box 3000, Boulder, CO 80307, United States
}
\altaffiltext{8}{
The Earth Simulator Center, Japan Agency for Marine-Earth Science and Technology, 3173-25 Showa-machi, Kanazawa-ku, Yokohama 236-0001, Japan
}
\altaffiltext{9}{
Space Sciences Laboratory, University of California, Berkeley, 7 Gauss Way, Berkeley, CA 84720, United States
}

\email{\myemail}


\begin{abstract}
We present a multi-wavelength observation of a solar flare occurring on 2006 December 13 with {\Hinode}, {\RHESSI}, and the Nobeyama Radio Observatory, to study the electron acceleration site and mechanism. 
The Solar Optical Telescope (SOT) on board {\Hinode} observed elongated flare ribbons, and {\RHESSI} observed double-footpoint hard X-ray (HXR) sources appearing in part of the ribbons.
 A photospheric vector magnetogram obtained from SOT reveals that the HXR sources are located at the region where horizontal magnetic fields change the direction. The region is interpreted as the footpoint of  magnetic separatrix. 
Microwave images taken with the Nobeyama Radioheliograph show a loop structure connecting the HXR sources.
The brighter parts of the microwave intensity are located between the top and footpoints of the loop.
We consider these observations as an evidence of the electron acceleration near the magnetic separatrix and injection parallel to the field line.
\end{abstract}


\section{Introduction}\label{sec:introduction}
It is widely recognized that a large amount of particles are accelerated to nonthermal levels in association with a solar flare. Electrons are accelerated to several tens of keV to MeV, and ions to several tens of MeV to GeV. A mechanism for accelerating particles is, however, not understood well. 
Nonthermal electrons are observed mainly through hard X-rays (HXRs), microwaves, and occasionally continuum $\gamma$-rays, whereas ions are observed through line $\gamma$-rays and ground level neutrons. These observations provide the most important information about the particle acceleration. It is also important for better understanding to obtain detailed information about global electromagnetic fields in which particles are accelerated.

Multi-wavelength observations have revealed that the evolution of the global magnetic fields in the flaring region significantly influences the dynamics of electrons.
 In many flares a plasma ejection, known as a plasmoid and coronal mass ejection (CME), is observed, as a consequence of magnetic reconnection \citep[e.g.,][]{1995ApJ...451L..83S}. The speed of the ejected plasma is not necessarily constant, which would reflect the non-stationarity of the magnetic reconnection process \citep[e.g.,][]{1997ApJ...487..437M}.
\cite{1998ApJ...499..934O} measured an acceleration rate of the plasmoid in the 1992 October 5 flare, and found that the acceleration phase coincides with the HXR impulsive phase. 
\cite{2008ApJ...673L..95T} confirmed these results using CME observations.
These indicate that the particle acceleration is related to the evolution of magnetic reconnection.

Flares frequently show motions of the footpoint and ribbons seen in HXR and H$\alpha$, which is thought to be a chromospheric response of the progress of magnetic reconnection \cite[e.g.,][]{2003ApJ...595L.103K,2005AdSpR..35.1707K}.
The convection electric field, estimated from their apparent motions and a photospheric magnetic field strength, is utilized for a measure of the power of magnetic reconnection \citep[e.g.,][]{1995SoPh..159..275L,2002GeoRL..29u..10I}.
By comparing the amplitude of the electric field with the corresponding HXR emission, it is found that stronger HXRs with harder energy spectrum are emitted at the site with stronger electric fields \citep{2002ApJ...565.1335Q,2004ApJ...611..557A,2007ApJ...654..665T,2007ApJ...664L.127J}. This suggests a physical relationship between the particle acceleration and the electric field generated by magnetic reconnection.

The {\Hinode} satellite \citep{2007SoPh..243....3K}, launched in 2006 September, boards three different types of scientific instruments with wavelengths ranging from optical to soft X-ray (SXR) bands, to study the dynamics of plasma and magnetic fields at the photosphere, chromosphere, and corona. 
On 2006 December 13, the first GOES X-class flare was observed by {\Hinode}.
The Solar Optical Telescope \citep[SOT;][]{2008SoPh..249..167T} successfully obtained the photospheric vector magnetograms before and after the flare \citep{2007PASJ...59S.779K,2008ApJ...682L.133Z}, allowing us to study the evolution of the magnetic fields during the course of the flare. These data are utilized for the Non-Linear Force-Free (NLFF) extrapolation of coronal magnetic fields \citep{2008ApJ...675.1637S,2008ApJ...676L..81J,2008ApJ...679.1629G,2008ASPC..397..110I}.
Dynamic motion of flare ribbons is seen in the high cadence Ca {\scriptsize II} H 3968 {\AA} and G band 4305 {\AA} images \citep{2007PASJ...59S.807I,2008ApJ...672L..73J}.
 The X-ray Telescope \citep[XRT;][]{2007SoPh..243...63G} observed the evolution of sheared coronal magnetic fields \citep{2007PASJ...59S.785S}. The EUV Imaging Spectrometer \citep[EIS;][]{2007SoPh..243...19C} measured plasma flows \citep{2007PASJ...59S.793I,2008ApJ...685..622A} and turbulent motions \citep{2008ApJ...679L.155I}.

In addition to these findings by {\Hinode}, HXR and microwave emissions from nonthermal electrons are observed with the {\it Reuven Ramaty High Energy Solar Spectroscopic Imager} \citep[{\RHESSI};][]{2002SoPh..210....3L}, the Nobeyama Radio Polarimeters \citep[NoRP;][and references their in]{1985PASJ...37..163N} and the Nobeyama Radioheliograph \citep[NoRH;][]{1994PROCIEEE...82..705}.
This flare is of great interest also for the study of the particle acceleration because {\Hinode} provides the unprecedentedly high quality data of the three-dimensional fields in which particles are produced, propagate, and finally dissipate. 
We expect to obtain new insights into the electron acceleration process by comparing {\RHESSI} and NoRH images with {\Hinode} observations.
For technical reasons, both {\RHESSI} and NoRH have problems in imaging in this flare. Since we have succeeded in solving the problems, it is possible to discuss the acceleration site and mechanism of electrons from the {\RHESSI}, NoRH, and {\Hinode} images.

In this paper we present the spatial distribution of the nonthermal emissions and the magnetic fields in the 2006 December 13 flare by using {\RHESSI}, NoRH, and {\Hinode}, to determine the electron acceleration site and mechanism. Observation data sets and results are presented in {\S} \ref{sec:data} and {\S} \ref{sec:results}, respectively. Based on the results, we discuss the acceleration site and mechanism in {\S} \ref{sec:summary-discussion}.

\section{Data Sets}\label{sec:data}
\begin{figure}[htbp]
\centering
\plotone{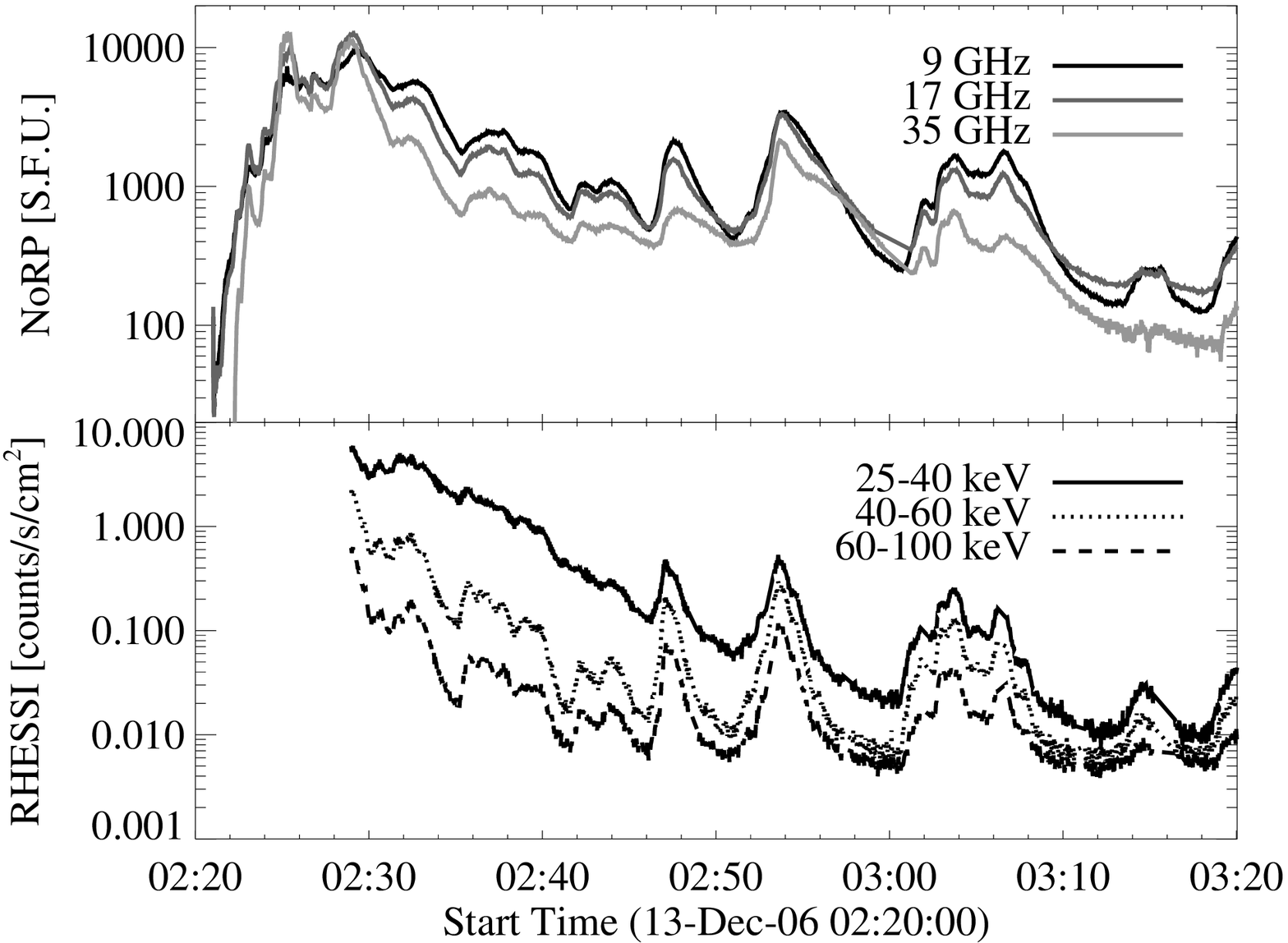}
\caption{Light curves for the 2006 December 13 flare. {\it Top}: Microwave light curves (S.F.U. = $10^{-19} \; {\rm erg/s/cm^2/Hz }$) taken with the NoRP 9.4 (black), 17 (gray), and 35 GHz (light gray) bands. {\it Bottom}: HXR light curves (in units of ${\rm counts/s/cm^2}$) taken at {\it RHESSI} in 25-40 (solid), 40-60 (dotted), and 60-100 keV (dashed) energy bands.}
\label{fig:ltc}
\end{figure}

The intense flare commenced at 02:20 UT, 2006 December 13, in NOAA active region 10930 (S$06^{\circ}$, W$22^{\circ}$). The GOES SXR level is X3.4 \citep[for the light curve, see][]{2008SoPh..247...53N}.
 Nonthermal HXRs and microwaves were observed for more than one hour. NoRP successfully detected quite intense microwave emissions from 1 to 80 GHz during the entire course of the flare. The NoRP microwave light curves in 9.4, 17, and 35 GHz during 02:20-03:20 UT are shown in the top panel of Figure \ref{fig:ltc}. The light curves consist of many spikes: the intense ones at 02:25 and 02:29 UT (impulsive phase), and the subsequent ones. The bottom panel shows the HXR light curves taken at {\RHESSI} in 25-40, 40-60, and 60-100 keV energy bands. The spikes of the HXR light curves correspond to the microwaves well. 
Unfortunately {\RHESSI} could not observe the first intense spike and the first half of the second intense spike seen in the microwave because of the satellite night.

{\RHESSI} image reconstruction \citep{2002SoPh..210...61H} only works for areas far enough away from the projected location of {\RHESSI}'s spin axis on the solar disk. The separation needs to be at least a few times larger than the spatial resolution of the subcollimators used for the imaging. For the flare discussed here, this is unfortunately not the case for most of the flare duration. However, we searched for the best possible time intervals (02:29-02:33 UT), and by using only fine subcollimators (numbers 3-5), we were able to reconstruct images with the usual quality\footnote{http://sprg.ssl.berkeley.edu/\texttt{\~}tohban/nuggets/?page=article\texttt{\&}article\texttt{\_}id=57}. A spatial resolution of {\RHESSI} images is $\sim 10^{''}$.

NoRH, a two-dimensional radio interferometer, obtains the correlation data, which are synthesized to make full Sun intensity images.
In a usual image synthesis, the procedure calibrates the position and intensity of the sources by using the quiet Sun (brightness temperature $10^{4} \; {\rm K}$) as a calibrator. This calibration method fails if quite intense ($\gsim 10^{7} \; {\rm K}$) sources appear. Unfortunately the NoRH data during 02:24-02:34 UT fail to be calibrated by the usual method. 
Thus, we apply another calibration method to determine the source position as follows; Two images are produced separately from the same correlation data. One is synthesized by using the calibration data obtained at the other time when the flare already rose intensively but the usual procedure still ran well. We consider that phases of the NoRH signals were stable during the flare activity so that the calibration data at the other time gives the absolute source position in the field of view of NoRH. The other is synthesized by using the calibration data obtained at each time without evaluating the absolute source position. The brightness distribution of the source in this image is more accurate than that in the former image. By combination of these two, the brightness distribution of the source with the absolute position is obtained. The calibration of the source intensity requires another procedure. Therefore we use the NoRH images for a morphological study only. A spatial resolution of NoRH images is $\sim 10^{''}$, and an accuracy of the absolute position in the field of view is better than the pixel size of the images $(\sim 2.5^{''})$.


We use the Ca {\scriptsize II} H and G band images taken by the Broadband Filter Imager (BFI) of SOT by 2 minutes cadence, and also use the photospheric vector magnetogram  obtained from the Spectro-Polarimeter (SP) of SOT at 20:30 UT, 2006 December 12, that is studied in \cite{2007PASJ...59S.779K}. 
The $180^{\circ}$ azimuth ambiguity of horizontal magnetic fields is removed by the AZAM utility \citep{1995ApJ...446..877L}.
However, the $180^{\circ}$ azimuth ambiguity can not be completely removed, particularly for regions with a complicated magnetic field distribution. In this paper we discuss only a large-scale horizontal magnetic field structure comparable to the spatial resolutions of {\RHESSI} and NoRH ($\sim 10^{''}$).
We also use the SXR images taken by XRT with Be-thin filter with a cadence of 1 minute. Details on the XRT observation are presented in \cite{2007PASJ...59S.785S}.

Since the field of view of SOT is smaller than the solar disk and XRT observed the flare with the partial-frame mode, co-alignments between these images and those taken with other instruments are necessary for our study. Using images taken with the {\it Solar and Heliospheric Observatory} ({\SOHO}) and {\it Transition Region and Coronal Explorer} ({\TRACE}), we co-align them as follows. First, we cross-correlate the longitudinal magnetogram taken by the Narrowband Filter Imager (NFI) of SOT with the {\TRACE} white light image, and the XRT image with the {\TRACE} EUV image. {\TRACE} image offsets of the different wavelength channels are determined \citep{1999SoPh..187..229H}, but the pointing data has an error as much as $10^{''}$. Then we further cross-correlate the {\TRACE} white light image with the continuum image taken with the Michelson Doppler Imager (MDI) on board {\SOHO}. 
The pointing data of {\SOHO}/MDI, {\RHESSI}, and NoRH are thought to be determined well.
Through these cross-correlations we compare the {\Hinode} images with the {\RHESSI} and NoRH. An accuracy of the co-alignments between these images is a few arcseconds, which is smaller than the spatial resolutions of {\RHESSI} and NoRH ($\sim 10^{''}$).

 
\section{Results}\label{sec:results}



\begin{figure}[htbp]
\centering
\plotone{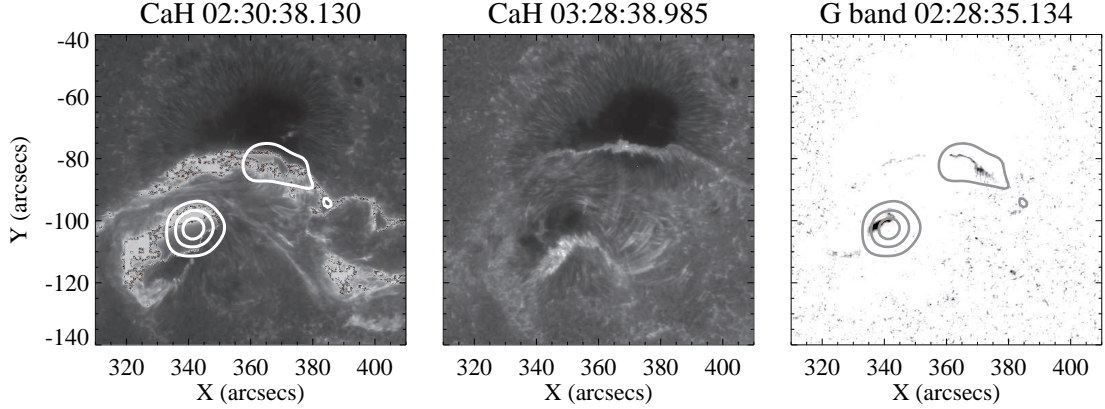}
\caption{Spatial distribution of the {\RHESSI} 35-100 keV HXRs (contours) at 02:29:04 UT on the SOT/BFI images, in the 2006 December 13 flare. Contour levels are 20, 50, and 80\% of the peak intensity. The left and middle panels show the Ca {\scriptsize II} H images at 02:30:38 UT and at 03:28:38 UT, respectively. The black spots in the left image are saturated data. The right panel shows the G band image at 02:28:35 UT. This image is drawn after removing the static features of the photosphere, to emphasize the flare-associated emissions.}
\label{fig:bfi_hsi_img}
\end{figure}

In Figure \ref{fig:bfi_hsi_img} we show the SOT/BFI images during the flare, and the {\RHESSI} 35-100 keV HXR image during 02:28:58-02:29:10 UT.
The Ca {\scriptsize II} H image shows the great flare ribbons elongated in east-west direction during the impulsive phase (left panel). 
They evolve toward north-south and separate from each other (middle panel). 
On the other hand, the HXR sources appear in part of the ribbons.
As shown later (Figs. \ref{fig:mag_hsi_img} and \ref{fig:xrt_hsi_norh_imgs}), they are magnetically-conjugate double-footpoint sources of a flare loop seen in SXRs and microwaves, originating from the electron precipitation.
 We confirm from the right panel of Figure \ref{fig:bfi_hsi_img} that they spatially coincide with the sites of the enhancement seen in the G band image.
This supports the interpretation that nonthermal electrons are the most powerful candidate for the primal energy source of the G band enhancement \citep{2007PASJ...59S.807I}.

In Figure \ref{fig:mag_hsi_img}, we show the vector magnetogram taken with SOT/SP (left, middle, and right panels for $B_x$, $B_y$, and $B_z$) and the {\RHESSI} 35-100 keV image (same as in Figure \ref{fig:bfi_hsi_img}). The magnetogram image is differentially rotated to the {\RHESSI} observation time. As can be seen from the longitudinal magnetogram (right panel), the HXR sources are certainly magnetic conjugates.

From the vector magnetogram, we can not identify a correlation between the east-west ribbon distribution (denoted as white dashed lines) and the horizontal magnetic fields (left and middle panels). 
On the other hand, the distribution of the HXR sources is closely related to the photospheric magnetic field distribution. 
The longitudinal magnetogram (right panel) shows that the HXR sources are located at the regions with the strongest magnetic field along the ribbons \citep[e.g.,][]{2002ApJ...578L..91A,2007ApJ...654..665T}.
We further find from the horizontal vector magnetogram that they are located only at which the direction of the horizontal magnetic fields changes remarkably.
To see this relation clearly, we measure the magnetic field components and the HXR intensities perpendicular to the southern and northern ribbons (denoted as black dashed lines in the middle panel), which are presented in the left and right panels of Figure \ref{fig:val_perp_ribbon}, respectively.
The relation is evident particularly in the northern ribbon (right panel). 
In the southeastern side of the HXR source, the horizontal magnetic field is directed southwestward ($B_x \sim 1000 \;{\rm G}$ and $B_y \sim -500 \; {\rm G}$). Around the HXR peak position the direction becomes almost northward ($B_x \sim 0$ and $B_y \sim 1500 \; {\rm G}$), and at last the magnetic field in the northwestern side is directed northeastward ($B_x \sim -2000 \; {\rm G}$ and $B_y \sim 1500 \; {\rm G}$). 
From the left panel, we confirm that the $B_y$ component changes the sign along the direction perpendicular to the southern ribbon: $B_y \sim 500 \; {\rm G}$ in the northeastern side and $-1500 \; {\rm G}$ in the southwestern side.
In addition, we can see from the left panel of Figure \ref{fig:mag_hsi_img} that the $B_x$ component changes the sign along the direction parallel to the ribbon.
Both the $B_x$ and $B_y$ components change the sign through the HXR source positions.
We consider that these large-scale variations of the horizontal magnetic fields are valid within the $180^{\circ}$ azimuth ambiguity.


In Figure \ref{fig:xrt_hsi_norh_imgs} we present the time sequence of the SXR images taken with XRT and the NoRH 34 GHz images during the flare.
The SXR images show two components. One is a highly sheared component lying in the east, which can be seen before (left panel) and during (middle panel) the impulsive phase. After the impulsive phase (right panel) this component is invisible, probably be relaxed through the flare. 
Another is a cusp-shaped loop connecting the HXR sources, located in the west (middle panel). This is thought to be formed by magnetic reconnection \citep{1992PASJ...44L..63T}. As time goes by, the loop evolves southwestward and forms an arcade structure (right panel).
The time evolution of the SXR loops is studied in detail by \cite{2007PASJ...59S.785S}. 

The microwave source also shows a loop structure connecting the HXR sources (middle panel).
We consider that the HXR- and microwave-emitting electrons are from the same population because their spatial distributions as well as light curves (Fig. \ref{fig:ltc}) are clearly related. 
Along the loop the microwave has two bright parts at $[350^{''},-110^{''}]$ and $[375^{''}, -100^{''}]$.
In Figure \ref{fig:hsi_norh_pol} we overlay the degree of the 17 GHz circular polarization image on the 34 GHz image at 02:29:10 UT, which shows that the eastern/western part is right-/left-circularly polarized, respectively. 
From the spatial distribution of $B_z$ (right panel of Fig. \ref{fig:mag_hsi_img}), we consider that the eastern/western part has positive/negative magnetic polarity.
The relationship between the microwave and magnetic polarities is in agreement with a sense of {\it x}-mode {\gyros} radiation \citep{1985ARA&A..23..169D}.
The polarization inverts around $[365^{''}, -110^{''}]$, between the bright parts.
It is plausible to speculate the top of the microwave loop around there. 
The two brighter parts of the microwave intensity are located between the loop top and footpoints (called ``legs''). 
As time goes by, the loop evolves southwestward, similar to the SXR loops.

\section{Summary and Discussion} \label{sec:summary-discussion}

\begin{figure}[htbp]
\centering
\plotone{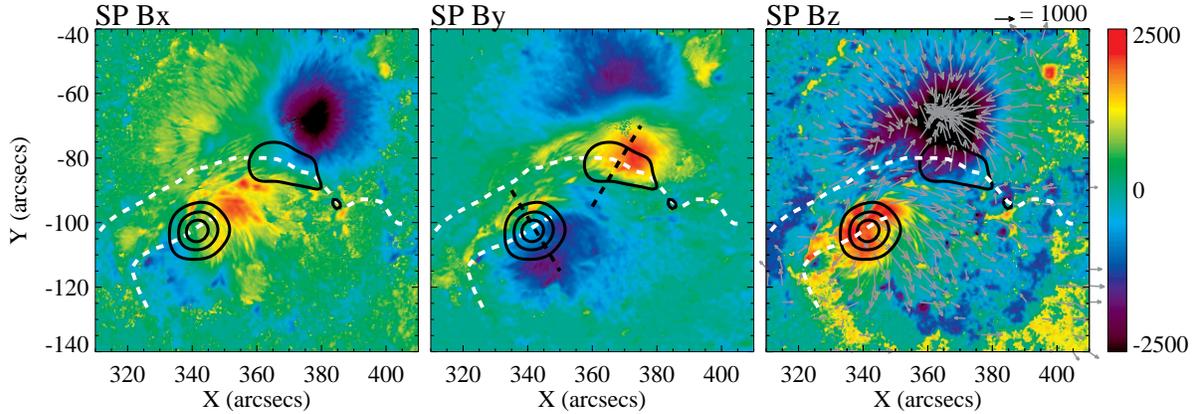}
\caption{The vector magnetogram taken with SOT/SP and the {\RHESSI} 35-100 keV image (black contours, same as in Fig. \ref{fig:bfi_hsi_img}) in the 2006 December 13 flare. Contour levels are 20, 50, and 80\% of the peak intensity. From left to right, the strengths of the horizontal magnetic fields $B_x$ and $B_y$, and the longitudinal magnetic field $B_z$ are shown. The color scale of the magnetic field strength is in units of Gauss. The sign of the strength represents the direction of the magnetic field vector. The horizontal magnetic field vectors are also indicated with gray arrows in the $B_z$ image. White dashed lines represent the position of the ribbons seen in the Ca {\scriptsize II} H image at 02:30:38 UT (left panel of Fig. \ref{fig:bfi_hsi_img}). Along the black dashed lines (shown in the middle panel) we measure the magnetic field components and the HXR intensity, which is presented in Figure \ref{fig:val_perp_ribbon}.}
\label{fig:mag_hsi_img}
\end{figure}
\begin{figure}[htbp]
\centering
\plotone{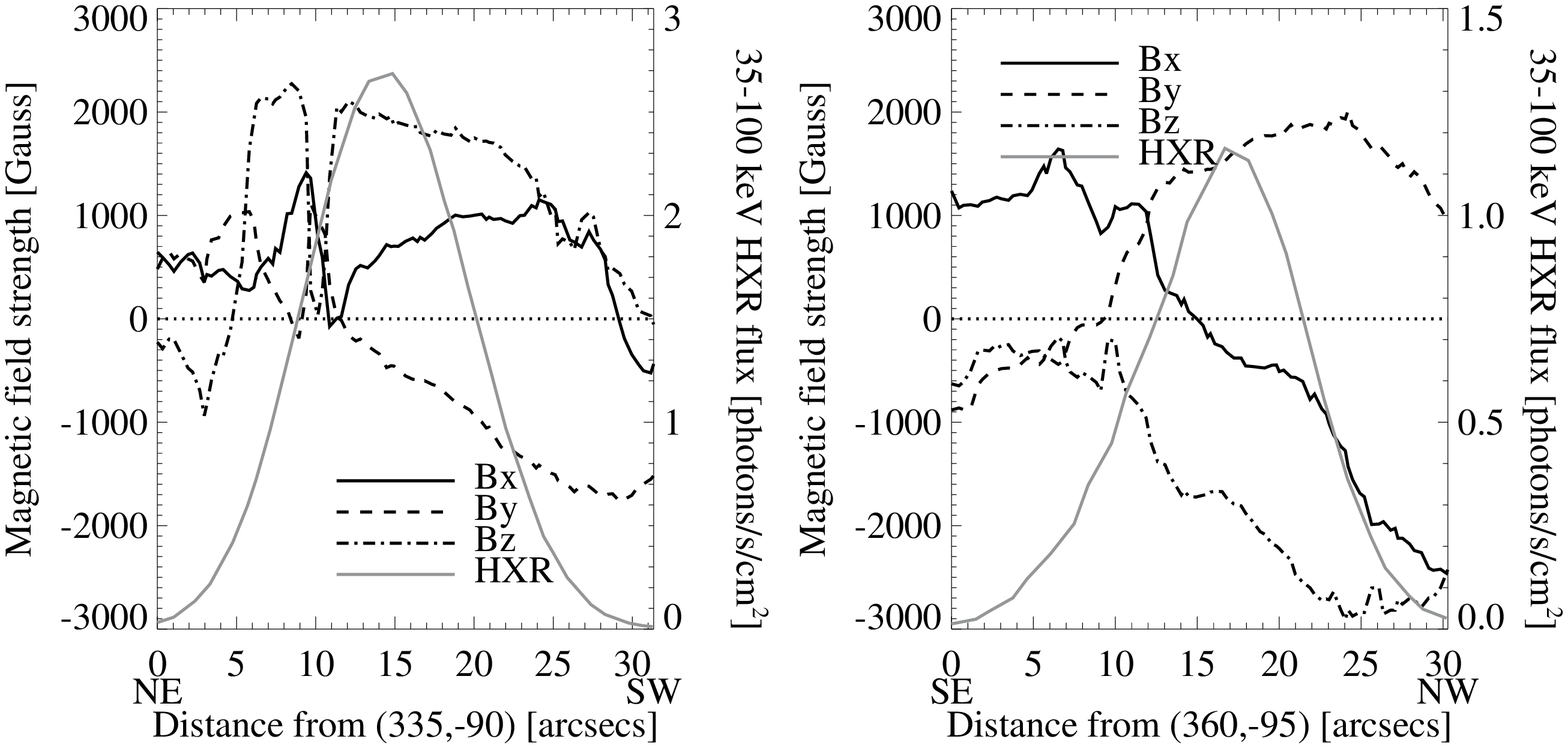}
\caption{The magnetic field components (solid, dashed, and dot-dashed lines for $B_x$, $B_y$, and $B_z$) and the HXR intensity (gray line) measured perpendicular to the southern (left) and northern (right) ribbons. Horizontal axes correspond to a spatial coordinate along the black dashed lines in the middle panel of Figure \ref{fig:mag_hsi_img}, from the northeast (NE, $[x,y]=[335^{''},-90^{''}]$) to the southwest (SW, $[x,y]=[350^{''},-115^{''}]$) ends for the left plot, and from the southeast (SE, $[x,y]=[360^{''},-95^{''}]$) to the northwest (NW, $[x,y]=[375^{''},-70^{''}]$) ends for the right plot. A dotted line shows that the magnetic field strength is zero.}
\label{fig:val_perp_ribbon}
\end{figure}

We present a multi-wavelength observation of the 2006 December 13 flare with {\RHESSI}, NoRP, NoRH, and {\Hinode}. Using the photospheric vector magnetogram from SOT/SP and the {\RHESSI} HXR image (Fig. \ref{fig:mag_hsi_img}), we find that the HXR sources are located at which the change of the direction of the horizontal magnetic fields is the most clearly seen. This implies that the sites of the HXR sources are the footpoints of magnetic separatrix. The implication is supported from the Non-Linear Force-Free (NLFF) coronal magnetic field configuration. \cite{2008ApJ...675.1637S} present the NLFF model of the coronal magnetic fields of this flare. We see from their result that the direction of magnetic field lines changes clearly around the HXR source positions. Hence we interpret our observation result as a direct evidence of the propagation of the accelerated electrons along the magnetic field line near the separatrix.

\begin{figure}[htbp]
\centering
\plotone{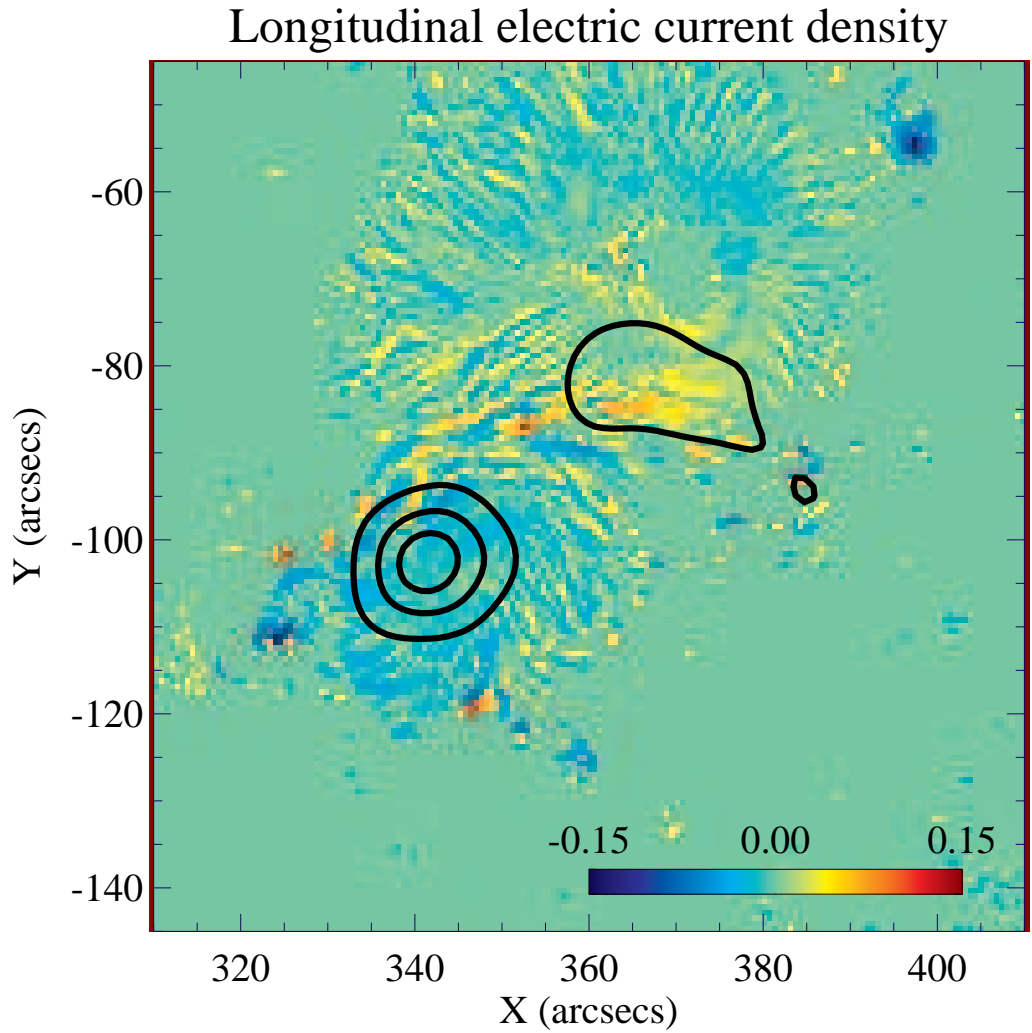}
\caption{Spatial distribution of the longitudinal electric current density, calculated from the horizontal magnetic fields taken by SOT/SP (left and middle panels of Fig. \ref{fig:mag_hsi_img}). The color scale is in units of ${\rm ampere/m^2}$. The {\RHESSI} 35-100 keV image (same as in Fig. \ref{fig:bfi_hsi_img}) is overlaid as black contours. Contour levels are 20, 50, and 80\% of the peak intensity.}
\label{fig:sp_jz_hsi}
\end{figure}

\cite{1993ApJ...411..362C}, \cite{1993ApJ...411..370L}, and \cite{1993ApJ...411..378D} examined the spatial relationship between longitudinal electric currents and sites of the electron precipitation, by using photospheric vector magnetograms and H$\alpha$ line profiles obtained from ground-based instruments. They found that the precipitation sites are located at the edges of longitudinal current channels, rather than at longitudinal current maxima. In Figure \ref{fig:sp_jz_hsi} we compare the HXR image (black contours) with the spatial distribution of the longitudinal electric current density, calculated from the horizontal magnetic fields (left and middle panels of Fig. \ref{fig:mag_hsi_img}) with Amp\`{e}re's Law. We can see the strong current channel laying around [350'', -90''] and the HXR sources at the edges of the channel. This is consistent with the previous finding. 

\begin{figure}[htbp]
\centering
\plotone{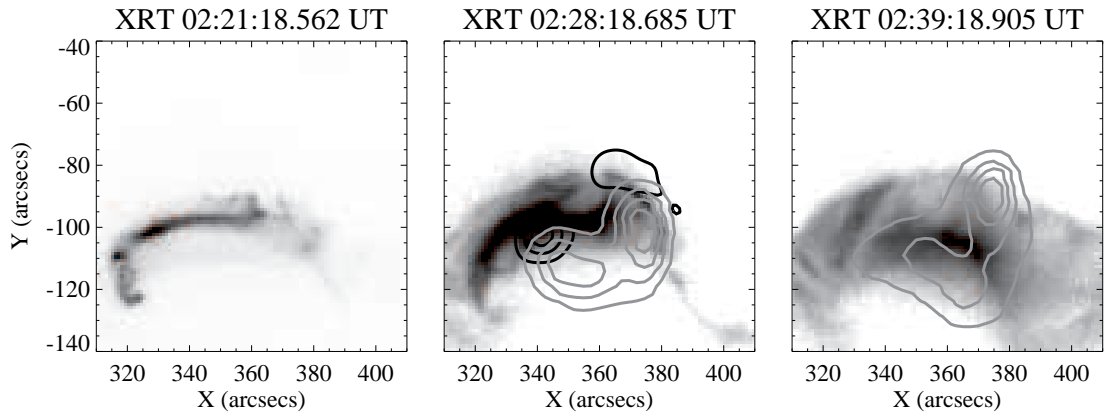}
\caption{Time sequence of the SXR negative images taken with XRT and the NoRH 34 GHz images (gray contours) in the 2006 December 13 flare. The XRT images are taken at 02:21:18 UT (left), 02:28:18 UT (middle), and 02:39:18 UT (right). The NoRH images are taken at 02:29:10 UT (middle) and 02:40:00 UT (right). Contour levels of the NoRH images are 20, 40, 60, and 80\% of the peak intensity. The {\RHESSI} 35-100 keV image at 02:29:04 UT is overlaid as black contours (same as in Fig. \ref{fig:bfi_hsi_img}). Contour levels of the {\RHESSI} image are 20, 50, and 80\% of the peak intensity.}
\label{fig:xrt_hsi_norh_imgs}
\end{figure}

\begin{figure}[htbp]
\centering
\plotone{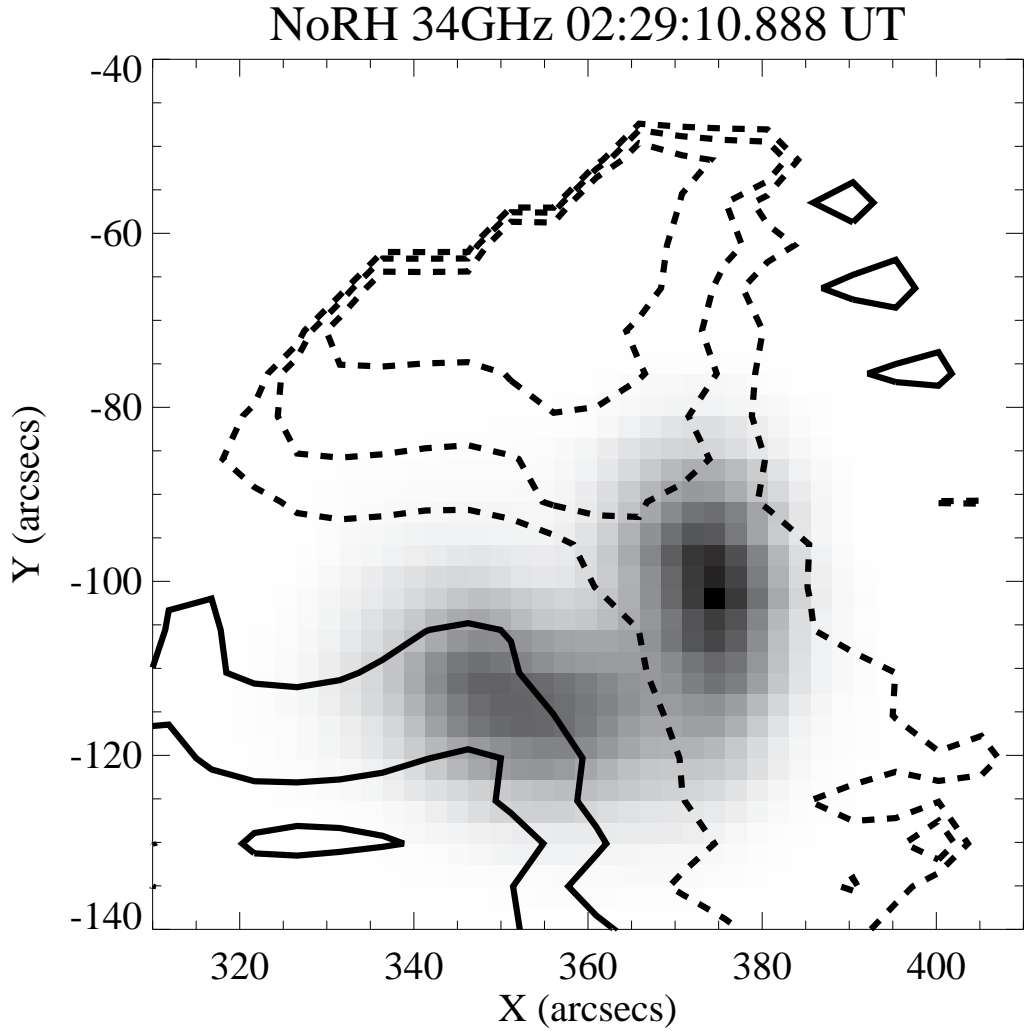}
\caption{Spatial distribution of the degree of right (thick contours) and left (dashed contours) circular polarization at NoRH 17 GHz, overlaid on the 34 GHz negative images, at 02:29:10 UT in the 2006 December 13 flare. Contour levels are 5, 10, and 20 \%.}
\label{fig:hsi_norh_pol}
\end{figure}

The microwave source shows the loop structure connecting the HXR sources (middle panel of Fig. \ref{fig:xrt_hsi_norh_imgs}). There are the brighter parts of the microwave intensity at the two legs. The eastern part with positive magnetic polarity is right-circularly polarized and vice verse (Fig. \ref{fig:hsi_norh_pol}), meaning that the microwaves are {\gyros} radiation by the accelerated electrons. 

The location of the microwave loop seems to be slightly different from that of the SXR loop: the microwave loop is located in the southwestern side of the SXR loop, or, is located higher than the SXR loop if the projection effect is significant (middle panel of Fig. \ref{fig:xrt_hsi_norh_imgs}). This displacement might be explained in terms of the temporal difference of the formation of these loops. The microwave loop represents the distribution of nonthermal electrons that would be produced in newly-reconnected magnetic field lines. On the other hand, the SXR loop represents the distribution of a thermal plasma from the chromosphere via the chromospheric evaporation, which is caused by the dissipation of thermal and nonthermal energies produced by magnetic reconnection. Therefore the SXR loop corresponds to the magnetic field lines that is reconnected earlier. This interpretation can explain the displacement because the loops evolved southwestward (right panel). Even if the displacement is mainly due to the projection effect, it is also consistent with the magnetic reconnection model that a later loop is formed higher in the corona \citep{2001ApJ...549.1160Y}.
 
From the microwave distribution and the configuration of the magnetic field, we discuss the pitch-angle distribution of the electrons injected into the loop.
Based on the magnetic reconnection model, we assume that the accelerated electrons with a mean pitch angle $\alpha = \alpha_{0}$ are injected into the loop top where the magnetic field strength is locally minimum, $B=B_0$. 
The {\gyros} radiation is generated primarily by electrons with pitch angles close to the viewing angle $\theta$ with respect to the magnetic field line at the radiation site \citep{1981ApJ...251..727P,1990ApJ...354..735L}.
Assuming the conservation of the magnetic moment, $\sin^2 \alpha/B = {\rm constant}$, we obtain
\begin{eqnarray}
\sin^2 \alpha_0 = \frac{B_0}{B_{\rm rad}} \sin^2 \alpha_{\rm rad} \sim \frac{B_0}{B_{\rm rad}} \sin^2 \theta_{\rm rad}, \label{eq:3}
\end{eqnarray}
where the variables with subscript ``rad'' are measured at the radiation site.
 
From the NLFF extrapolation of the coronal magnetic fields based on the photospheric vector magnetogram \citep{2008ASPC..397..110I}, we find that $\sin \theta$ at the loop top is close to unity and decreases toward the footpoints.
When the microwave peaks at the legs ($B_{\rm rad} > B_0$ and $\sin \theta_{\rm rad} < 1$), equation (\ref{eq:3}) gives $\sin \alpha_{0} < 1$, meaning the electron injection parallel to the field line.
A crude estimation of the magnetic field strength and the viewing angle at the leg, $B_0/B_{\rm rad} \sim 0.5 \; {\rm and} \; \theta_{\rm rad} \sim 45^{\circ}$, gives $\alpha_0 \sim 30^{\circ}$.
\cite{2008ApJ...686..701M} demonstrated a numerical simulation of the electron propagation and the resulting {\gyros} radiation along a magnetic loop. The simulation confirms that the {\gyros} intensity peaks at the leg, when the electrons with small pitch angles are injected into the loop located at the disk center. 
We interpret the spatial distribution of the microwave presented in the middle panel of Figure \ref{fig:xrt_hsi_norh_imgs} as a result of the electron injection parallel rather than perpendicular to the magnetic field line.



The electron pitch-angle distribution can be discussed also from the temporal evolution of the spectrum of nonthermal emissions \citep{2008ApJ...673..598M}. When a larger fraction of electrons are injected into a loop with large pitch angles, the spectrum of microwaves, emitted by electrons trapped in the loop, are more likely to show the ``{\it Soft-Hard-Harder}'' evolution \citep[e.g.,][]{2000ApJ...545.1116S}. On the other hand, the spectrum of HXRs, emitted by precipitating electrons into the footpoint, frequently shows the ``{\it Soft-Hard-Soft}'' evolution \citep[e.g.,][]{2004A&A...426.1093G}. \cite{2008SoPh..247...53N} examined the temporal evolution of the microwave spectrum in the flare with NoRP. The microwave during 02:29-02:32 UT, which we focus on in this paper, shows the ``{\it Soft-Hard-Soft}'' spectral evolution, same as the HXR from {\RHESSI}. This indicates that the microwave-emitting electrons are distributed in parallel rather than perpendicular to the magnetic field line. His result is consistent with our interpretation of the microwave image.


Based on the results and discussions, we conclude that the nonthermal electrons in the 2006 December 13 flare propagate preferably along the magnetic field line near the magnetic separatrix.
It is expected from the photospheric horizontal magnetic field distribution that magnetic field lines change the direction very steeply at this separatrix.
Subsequently the electrons are injected parallel to the field line.
This conclusion suggests a scenario for an effective acceleration of electrons toward the parallel direction taking place near the separatrix.

Let us discuss a probable acceleration mechanism. From both the XRT images and the NLFF extrapolation of the coronal magnetic fields, we find the magnetic transition from a NLFF to potential-like state (\citealt{2008ASPC..397..110I}; {Inoue et al. 2009, in preparation}). Such a topological change of magnetic fields induces electric fields. When the magnetic field line has finite curvature, electrons move against the inductive electric field due to the curvature drift, and then the velocity of the curvature drift increases.
Since the curvature drift velocity is proportional to the parallel kinetic energy, this mechanism accelerates particles parallel to the magnetic field line \citep[e.g.,][]{1963RvGSP...1..283N}. 
The rate of the acceleration due to the curvature drift is written as
 \begin{eqnarray}
\dif{K}{t} \simeq \frac{1}{R_c} \frac{cE}{B} K, \label{eq:1}
\end{eqnarray}
where $K$ is the parallel kinetic energy of a particle, $R_c$ is the curvature radius of the magnetic field line, $E$ is the electric field strength, and $c$ is the speed of light, respectively. 
The equation gives
\begin{eqnarray}
K_f \simeq K_i \exp\left[\frac{1}{R_c} \frac{cE}{B} \tau_{\rm acc}\right], \label{eq:4}
\end{eqnarray}
where $\tau_{\rm acc}$ is the acceleration time, and subscripts $i$ and $f$ mean the initial and final states.

We consider that magnetic reconnection yields a topological change of magnetic fields, and then induces electric fields. Following the magnetic reconnection model in an ideal magnetohydrodynamic (MHD) regime, we estimate $cE/B = v_A \sim 2000 \; {\rm km \; s^{-1}}$, the {\Alfven} velocity. Observationally, a typical time scale for the electron acceleration is 1 s \citep[e.g.,][]{1983ApJ...265L..99K,1984ApJ...287L.105K}. Hence we obtain from equation (\ref{eq:4}) that the 2 keV electrons are accelerated to 100 keV (HXR energy range) with $R_c \sim 500 \; {\rm km}$.
Stronger curvature is required for accelerating particles to higher energy.
We expect that a fine structure of the magnetic field line is realized particularly near the separatrix, at which the topological change of the magnetic fields from an open to close state takes place through magnetic reconnection.

It is thought that the curvature radius can become small as much as the width of the current sheet at the center of the reconnection system. In the two-dimensional magnetic reconnection model, the width $d$ is approximated from the mass conservation of the incompressive fluid, $d \sim (v_{\rm in}/v_A) L = M_A L$. Here $v_{\rm in}$ is the velocity of the magnetic reconnection inflow, $M_A$ is the reconnection rate, and $L$ is the length of the current sheet.
The reconnection rate in solar flares has been measured by many authors \citep[e.g.,][]{1996ApJ...472..864D,2001ApJ...546L..69Y,2002ApJ...566..528I,2005ApJ...622.1251L,2005ApJ...632.1184I,2006ApJ...637.1122N,2006ApJ...647..654N}, giving $M_A \sim 0.01$. Evaluating the length $L$ as a typical size of flares, $L \sim 10000 \; {\rm km}$, we obtain $R_c \gsim d \sim 100 \; {\rm km}$. This scale is sufficient to accelerate electrons due to the curvature drift, suggested before.


In addition, a slow-mode shock is expected to be attached near the separatrix \citep{1964psf..conf..425P}, which is observed in association with the magnetic reconnection events in the Earth magnetotail \citep{1995JGR...10023567S} as well as solar flares \citep{1996ApJ...456..840T}. At the front of the shock where magnetic field lines have strong curvature, particles may be accelerated further \citep{1997JGR...10222301S}.

A mechanism to trap particles in the acceleration site is needed for an efficient acceleration. 
This is problematic especially for the case of parallel acceleration of electrons because accelerated electrons are more likely to escape from the site.
Linear {\Alfven} waves, which are expected to be everywhere in the corona, do not work on the electron trapping.
Electrons could be trapped by the magnetic mirror force from large-amplitude {\Alfven} waves or by the resonance with high-frequency whistler waves, but the origin of these waves is unclear. This problem still remains open.



{\Hinode} provides the unprecedentedly high quality data of the photospheric vector magnetogram. 
Using these data as initial and boundary conditions, a data-driven MHD simulation is performed to describe the evolution of the plasma and magnetic fields in the 2006 December 13 flare \citep{2008AGUFMSH52A..06K}. These simulation data are of great importance also for the study of particle acceleration. A test-particle simulation in the fields obtained from the data-driven simulation is one of useful means to further understand the observations. Now we have a great opportunity to reveal the particle acceleration in solar flares theoretically and observationally.

\begin{acknowledgements}
The authors would like to sincerely thank K. Shibasaki for fruitful discussions and providing the NoRP and NoRH data. 
We also thank the anonymous referee for helpful comments and corrections to improve our manuscript.
{\Hinode} is a Japanese mission developed and launched by ISAS/JAXA with NAOJ as domestic partner and NASA and STFC (UK) as international partners. This work is supported partly by COE program of the University of Tokyo, ``Predictability of the Evolution and Variation of the Multi-scale Earth System: An integrated COE for Observational and Computational Earth Science'', and by the Grant-in-Aid for Creative Scientific Research of MEXT/Japan, the Basic Study of Space Weather Predication. 
\end{acknowledgements}

\bibliographystyle{apj}                                                       

\end{document}